\documentclass[aps,prl,onecolumn,showpacs,preprintnumbers,superscriptaddress]{revtex4}
\usepackage{bm,amsmath,amssymb,latexsym,mathrsfs,enumerate}
\usepackage[mathcal]{euscript}
\usepackage[hypertex]{hyperref}
\usepackage{graphicx}

\begin{document}

\title{NUTATIONAL TWO-DIMENSIONAL STRUCTURES IN MAGNETS}

\author{A. B. Borisov}
\email{Borisov@imp.uran.ru}
\affiliation{Institute of Metal Physics, Ural Division, Russian Academy of Sciences, Yekaterinburg 620041, Russia}

\author{F. N. Rybakov}
\email{F.N.Rybakov@gmail.com}
\affiliation{Institute of Metal Physics, Ural Division, Russian Academy of Sciences, Yekaterinburg 620041, Russia}

\date{19 October, 2007}
\begin{abstract}
New types of magnetic structures in the Heisenberg model are found. Analytical methods are
used to describe spiral structures, spiral vortex structures, and their interaction. Methods for obtaining
these structures in real systems, including nanomagnets, are discussed.
\end{abstract}

\pacs{05.45.Yv, 67.30.he}

\maketitle

\section{Introduction}

There has been great interest in vortices and other nonlinear
structures in low-dimension magnets over the last two
decades. Many investigators have noted the important role of
such structures in magnetic topological phase transitions
(see, for example, \cite{1} and \cite{2}). Now that a large class of
new quasi-one- and quasi-two-dimensional (2D) ferromagnets
(layered magnets, intercalated compounds, graphite, cuprate
planes in HTSCs \cite{3}), where the magnetic interaction
within crystallographic planes is much stronger than
the interaction between planes have been discovered and
synthesized, the theoretical description of nonlinear structures
and magnets is not only of academic interest.

Aside from different types of dynamical solitons, many
stationary and autowave structures of condensed media,
whose order parameter is a complex scalar field $\Psi$ just as in
the Heisenberg model ($\Psi=e^{i \Phi}cot(\theta)$ in the Heisenberg
model) and which are invariant with respect to global
changes at the phase $\Phi$, have now been studied. The most
popular and well studied equations of this form are the
Gross-Pitaevskii equation in the theory of superfluidity \cite{4}
\begin{equation}
{i\frac{\partial {{\Psi}}}{\partial t}={{\Delta}{\Psi}}+{{\lambda}{\Psi}}\pm {{\Psi}}|{{\Psi}}|^{{2}}}
\label{eq:eq1}
\end{equation}
and the complex Landau-Ginsburg equation \cite{5}
\begin{equation}
{i\frac{\partial {{\Psi}}}{\partial t}={{\Delta}{\Psi}}+{(1 + i b){\Psi}}\pm {(1 + i c){\Psi}}|{{\Psi}}|^{{2}}},
\label{eq:eq2}
\end{equation}
with real parameters $b$ and $c$ characterizing the linear and
nonlinear dispersion. For $\Psi=S_x+iS_y$ Eq. (\ref{eq:eq1}) also describes
on the basis of the Landau–Lifshitz equation the lowamplitude
dynamics of the spin system of a uniaxial ferromagnet
with a distinguished axis z. A nontrivial structure
described by Eq. (\ref{eq:eq1}) is the nonstationary vortex (for example,
the Pitaevskii vortex or magnetic vortex). In a polar
coordinates system the function $\Psi$ has the form $\Psi=\rho e^{i \Phi}$,
and the vortex solution is determined by substituting $\rho=\rho(r)$ and $\Phi=q \varphi$ into Eq. (\ref{eq:eq1}) and solving these equations
numerically with the boundary conditions $\rho(0)=0$, $\rho(\infty)=1$.
Two-dimensional vortices in isotropic ferromagnets and
magnets with uniaxial anisotropy were predicted a long time
ago (see references in the first monograph concerning this
problem \cite{6}, concerning magnetic solitons and magnetic structures
) and have been found recently in permalloy and nanomagnets
(magnetic nanodots); see, for example, \cite{7}.

Other nontrivial spatial structures observed in autooscillatory
active media are $N$-turn \emph{spiral waves}. They correspond
to solutions of Eq. (\ref{eq:eq2}) in the form $\rho=\rho(r)$, $\Phi=N \varphi+\omega t+f(r)$.
The phase of the spiral at large distances from the
center of the spiral is proportional to the distance $f(r\rightarrow\infty)=k r$, the frequency
$\omega=c+(b-c)^2 k^2$ is determined by the
characteristics of the auto-oscillatory system itself, $\rho(r)\rightarrow const$ as $r\rightarrow\infty$	, and the lines of constant phase are
Archimedean spirals. The wave amplitude decreases near the
center and vanishes as $r\rightarrow 0$. Numerical methods are used to
determine the form of the functions $\rho(r)$ and $f(r)$  \cite{8,9,10,11}. Spiral
structures are the richest class of spatial structures in active
media (spiral galaxies, mollusks, spiral waves in the
Belousov-Zhabotinskii reactions). Such media are characterized
by a continuous inflow of energy from a source to each
physically small element and dissipation of this energy, and
various stationary or time-dependent spatial structures,
which lie at the foundation of self-organization, are observed
to form in them under certain conditions.

It has been established experimentally \cite{12,13,14} that new
structures --- target-type guiding centers, spirals and spiral domains,
observed by means of the magneto-optic Kerr
effect --- are formed in thin magnetic films with strong perpendicular
easy-axis anisotropy under the action of harmonic
or pulsed magnetic fields. Static stability and strong nonlinearity
are characteristic experimentally observed features of
magnetic structures. They do not vanish after the magnetic
field is switched off --- the lifetimes of the targets and spiral
domains are several orders of magnitude greater than the
period of the magnetic field. This makes it possible to study
magnetic structures of the target and spiral-domain type as
magnetic defects which are excited by energy pumping and
relax to the thermodynamic equilibrium state over quite long
periods of time.

A detailed investigation of spiral vortex structures for the
$XY$ and Heisenberg models (since they are universal) is of
interest for investigating structures in nanomagnets, where
spiral structures have been found, as well as for possible
applications in the physics of liquid crystals and the quantum
Hall effect and for studying a number of biological systems
where self-organizing spiral structures have been found \cite{15,16}.

Very simple types of two-dimensional spiral vortex
structures have been obtained in analytic form for the
Heisenberg model of a ferromagnet \cite{17,18}. The most general
type of spiral structures which are formed by the main (exchange)
approximation are found in next section.  A wide class of
new, exact solutions of the corresponding equations is discussed,
the structure and interaction of spiral vortices are
investigated, and the possibility of obtaining such structures
experimentally is discussed.

\section{Two-dimensional spiral vortices}

We shall examine a model of an isotropic Heisenberg
ferromagnet with spin $S$ described by a Hamiltonian of the form
\begin{equation}
H=-\sum_{p,n}J_{pn}{\bf S}_{p}{\bf S}_{n},
\label{eq:eq3}
\end{equation}
where ${\bf S}_{p}$ is the spin operator at the site $p$ of a two- or three-dimensional
lattice and $\bf a$ is the distance between the nearest
neighbors with constant exchange interaction between them
$J_{pn}=J \delta_{n,p+{\bf a}}$ ($J>0$). The nonlinear differential equations describing
the dynamics of the model can be derived by examining
the diagonal matrix element of the equation of motion
of the $p$-spin operator ${S}^{+}_p ={S}^{x}_p +i {S}^{y}_p$:
\begin{equation}
-i \hbar \frac{d {S}^{+}_p}{d t}=[H,{S}^{+}_p]
\label{eq:eq4}
\end{equation}
in the representation of spin coherent states $|\Omega\rangle=\prod_p|\theta_p,\Phi_p\rangle$,
where $0\leq\theta_p\leq\pi$ and $0\leq\Phi_p<2\pi$ parameterize
the spin states on a sphere with unit radius \cite{20}. For a
Hamiltonian with bilinear interactions this results in a system
of equations for the classical variables $\theta_p$ and $\Phi_p$ parameterizing
the spin vector  ${\bf S}_{p}=S(sin \theta_p cos \Phi_p, sin \theta_p sin \Phi_p, cos \theta_p)$,
\begin{equation}
sin \theta_p \frac{\partial {\Phi_p}}{\partial t}=-\frac{S}{\hbar}\sum_n J_{np}sin \theta_p cos \theta_p cos(\Phi_p - \Phi_n)+ sin \theta_p \frac{S}{\hbar}\sum_n J_{np}cos \theta_n;
\label{eq:eq5}
\end{equation}
\begin{equation}
\frac{\partial {\theta_p}}{\partial t}=\frac{S}{\hbar}\sum_n J_{np}sin \theta_n sin(\Phi_n - \Phi_p),
\label{eq:eq6}
\end{equation}
here the index $n$ enumerates the nearest neighbors
of the spin under consideration. In the continuum limit in the
two-dimensional case we introduce the fields $\theta(x,y)$ and
$\Phi(x,y)$, which are defined in the $(x,y)$ plane. The equations
for the static solutions $\frac{\partial {\theta_p}}{\partial t}=\frac{\partial {\Phi_p}}{\partial t}=0$ can be obtained by
passing to the continual approximation in the equations for
spins on a discrete lattice
\begin{equation}
\begin{cases}
\Delta\theta=sin(\theta) cos(\theta) (\nabla\Phi) ^2,
\\
\nabla(sin(\theta)^2 \nabla\Phi)=0.
\end{cases}
\label{eq:eq7}
\end{equation}

It is shown in \cite{17,18}  that spiral structures exist
when the contour lines of the fields $\Phi(x,y)$ and $\theta(x,y)$ are
orthogonal ($\nabla\Phi \cdot \nabla\theta=0$) or parallel ($\nabla\Phi \propto  \nabla\theta$) to one another.
Here we shall examine the case where the derivative
fields $\Phi(x,y)$ are linear combinations of the derivatives of
the field $\theta$ with coefficients that depend on this field
\begin{equation}
\begin{cases}
\frac{\partial \Phi}{\partial x}=-F_1(\theta) \frac{\partial \theta}{\partial y}+F_2(\theta) \frac{\partial \theta}{\partial x},
\\
\frac{\partial \Phi}{\partial y}=F_1(\theta) \frac{\partial \theta}{\partial x}+F_2(\theta) \frac{\partial \theta}{\partial y},
\end{cases}
\label{eq:eq8}
\end{equation}
where, because the exchange interactions are invariant under
the rotation group, we assume the functions $F_1$ and $F_2$ to
depend only on the field $\theta(x,y)$. The compatibility condition
for the system (\ref{eq:eq8}) and (\ref{eq:eq7}) yields a closed system of three
nonlinear equations for the fields $\theta(x,y)$,  $F_1$, and  $F_2$. The
solution of two of them determines the fields  $F_1$ and  $F_2$ in
explicit form:
\begin{equation}
F_2=c_1 \frac{F_1}{sin(\theta)^2},\qquad F_1=\frac{2 sin(\theta)}{\sqrt{-4 c_1^2 - c_2 sin(\theta)^2 - sin(2 \theta)^2}}
\label{eq:eq9}
\end{equation}
with arbitrary constants $c_1$ and $c_2$, and the last equation has
the form
\begin{equation}
\Delta\theta=-(\nabla\theta)^2 \frac{F_1{'}}{F_1}.
\label{eq:eq10}
\end{equation}
Redefining the constants $c_1$ and $c_2$ and introducing the auxiliary
field $a(x,y)$, this equation reduces to the Laplace equation
\begin{equation}
\Delta a(x,y)=0,
\label{eq:eq11}
\end{equation}
where the field $a(x,y)$ determines the field $\theta(x,y)$ as follows:
\begin{equation}
cos[ \theta(x,y)]=c \cdot sn[a(x,y),k],\qquad (0<k<1).
\label{eq:eq12}
\end{equation}
Here $sn[a(x,y),k]$ is the Jacobi elliptic function (elliptic sine) with modulus $k$.

The solution of the system (\ref{eq:eq8}) becomes
\begin{equation}
\Phi(x,y)=- \frac{\sqrt{(1-c^2)(c^2-k^2)}}{c}\int_0^{a(x,y)} \frac{dX}{ (1-c^2 sn(X,k)^2)} + \Psi(x,y) .
\label{eq:eq13}
\end{equation}
where $\frac{k}{c}a(x,y)+i \Psi(x,y)$ is an analytic function of the
complex variable $z=x+i y$. As a result, the relations (\ref{eq:eq11}),(\ref{eq:eq12}),(\ref{eq:eq13})
give a new class of exact solutions of the equations  (\ref{eq:eq7}) which
is determined by the analytic function $\frac{k}{c}a(x,y)+i \Psi(x,y)$
and the two parameters $k$, $c$ whose ranges are
\begin{equation}
0\leq k \leq1, \qquad k\leq c \leq1.
\label{eq:eq14}
\end{equation}
The limits of the intervals of these parameters correspond to
known classes of solutions. In the limit $k\rightarrow0$ we obtain a
family of solutions which depend only on the harmonic function
$a(x,y)$
\begin{eqnarray*}
cos[ \theta(x,y)] & = & c \cdot sin[a(x,y)]
\\
\Phi(x,y) & = & -arctan(\sqrt{1-c^2}\cdot tan[a(x,y)])
\end{eqnarray*}
and were investigated in \cite{18}. For $c=1$ the substitution
(\ref{eq:eq8}) reduces to the Cauchy-Riemann equation for the analytic
function $\Omega=\Phi(x,y) + i k a(x,y)$ and
\begin{eqnarray*}
cos[ \theta(x,y)] & = & sn[a(x,y),k],\qquad (0<k<1)
\end{eqnarray*}
the corresponding class of solutions found in  \cite{17}. Finally,
for $c=1$, $k\rightarrow 1$, and
\begin{eqnarray*}
\Omega & = & \sum_{j=1}^n (Q_j) ln(x + i y - c_j),\qquad (Q_j\in Z)
\end{eqnarray*}
we represent the solutions as follows:
\begin{eqnarray*}
cot \left(\frac{\theta}{2}\right) e^{i \Phi}  & = & \prod_{j=1}^n {\left(\frac{x + i y - c_j}{A_j} \right) }^{Q_j}
\end{eqnarray*}
These solutions describe the structure and interaction of
instantons—magnetic vortices  \cite{20}.

Here we shall discuss the choice of $\frac{k}{c}a(x,y)+i \Psi(x,y)$
 in the form of a potential of vortex sources in hydrodynamics:
\begin{eqnarray*}
a(x,y) = \sum_{i=1}^n \alpha_i \cdot ln \left( \sqrt{(x-x_{0i})^2 + (y-y_{0i})^2}\right) + q_i \cdot arctan \left( \frac{y-y_{0i}}{x-x_{0i}}\right),
\end{eqnarray*}
\begin{equation}
\Psi(x,y) = \sum_{i=1}^n \left(- \frac{k}{c}\right) q_i \cdot ln \left( \sqrt{(x-x_{0i})^2 + (y-y_{0i})^2}\right) +  \left(\frac{k}{c}\right)  \alpha_i \cdot arctan \left( \frac{y-y_{0i}}{x-x_{0i}}\right),
\label{eq:eq15}
\end{equation}
with singularities at the points $(x_{0i},y_{0i})$ --- the centers of magnetic
defects. It follows from the fact that the magnetization
is single-valued, the form of the solution (\ref{eq:eq12}), and the symmetry
of the elliptic functions $sn(u,k)=sn(u+4K,k)$ and $sn(u,k)=sn(2K-u,k)$ (where $K=K(k)$ is the complete elliptic integral
of the first kind) that the changes in the fields $a$, $\Phi$ with one
revolution about a closed contour around the point $(x_{0i},y_{0i})$
must satisfy the conditions
\begin{equation}
\delta a=4 K N_i,\qquad \delta \Phi=2 \pi Q_i, \qquad (N_i, Q_i \in Z).
\label{eq:eq16}
\end{equation}
Hence follows immediately the macroscopic quantization of
the parameter $q_i$:
\begin{equation}
q_i = \frac{2 K N_i}{\pi}.
\label{eq:eq17}
\end{equation}
Taking account of the change of the field $\Phi(x,y)$ with such a
revolution, equal to
\begin{equation}
\delta\Phi(x,y)=- \frac{\sqrt{(1-c^2)(c^2-k^2)}}{c}\int_0^{\delta a(x,y)} \frac{dX}{ (1-c^2 sn(X,k)^2)} + \delta\Psi(x,y),
\label{eq:eq18}
\end{equation}
and the relation (\ref{eq:eq16}), we obtain a relation between the parameters
$\alpha_i$ and $Q_i$:
\begin{equation}
\alpha_i=\frac{c \pi Q_i + 2 \sqrt{1-c^2}\sqrt{c^2-k^2}N_i P(-c^2,k)}{k \pi},
\label{eq:eq19}
\end{equation}
where
\begin{equation}
P(-c^2,k)=\int_0^K \frac{dX}{(1 - c^2 sn(X,k)^2)}
\label{eq:eq20}
\end{equation}
is a complete elliptic integral of the third kind.

\begin{figure}
\includegraphics[width=\columnwidth, viewport=40 110 560 330,clip]{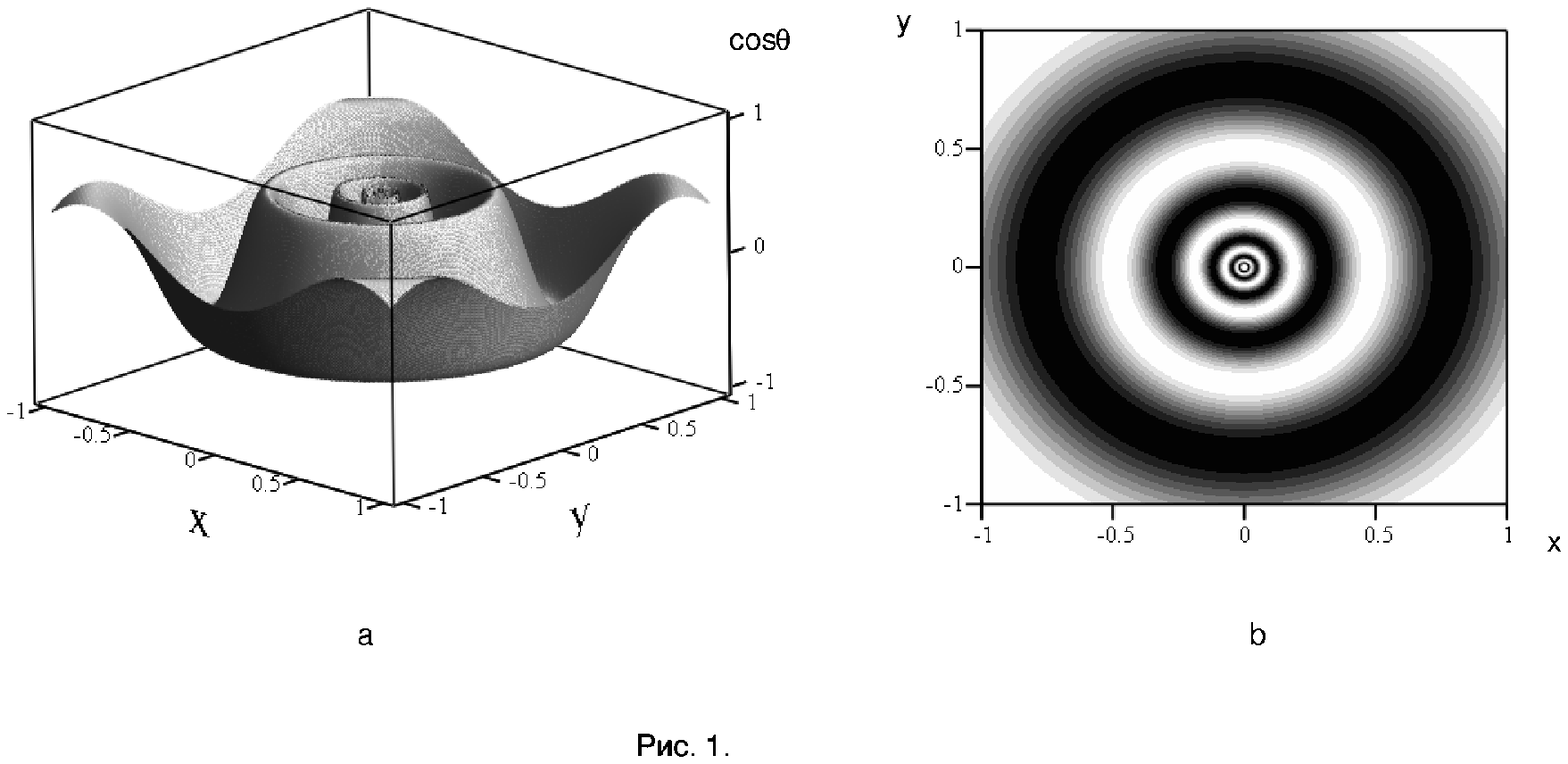}
\caption{Structure of the core of a vortical magnetic target ($Q=2$, $N=0$, $c=0.5$, $k=0.15$); plot of $cos(\theta)$ (a), regions with positive (white) and negative (black) values of $cos(\theta)$ (b).}
\label{F:fig-1}
\includegraphics[width=\columnwidth, viewport=40 110 560 340,clip]{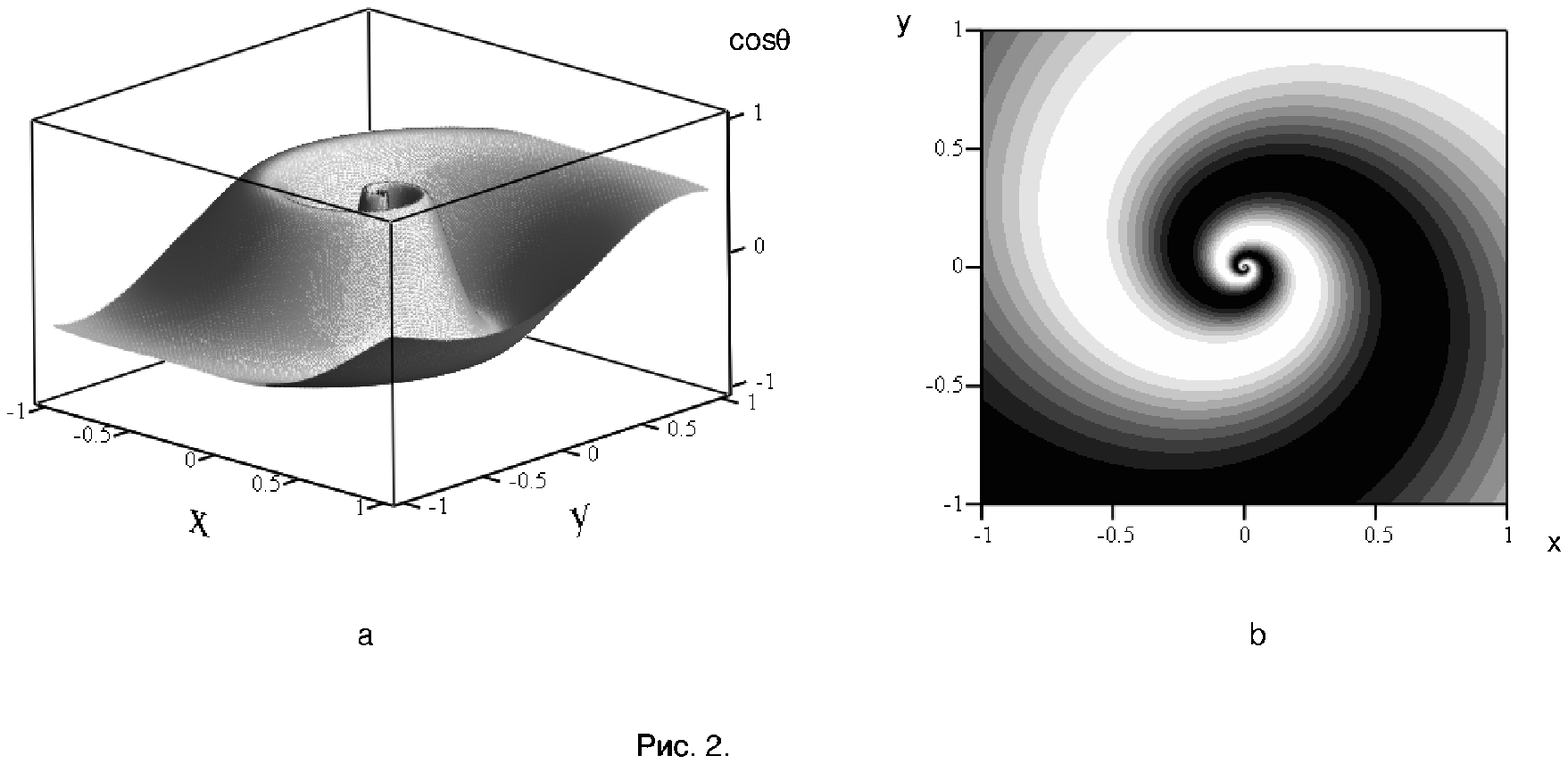}
\caption{Structure of the core of a one-turn spiral ($Q=0,N=1$, $c=0.5$, $k=0.15$).}
\label{F:fig-2}
\end{figure}

\begin{figure}
\includegraphics[width=\columnwidth, viewport=40 110 560 340,clip]{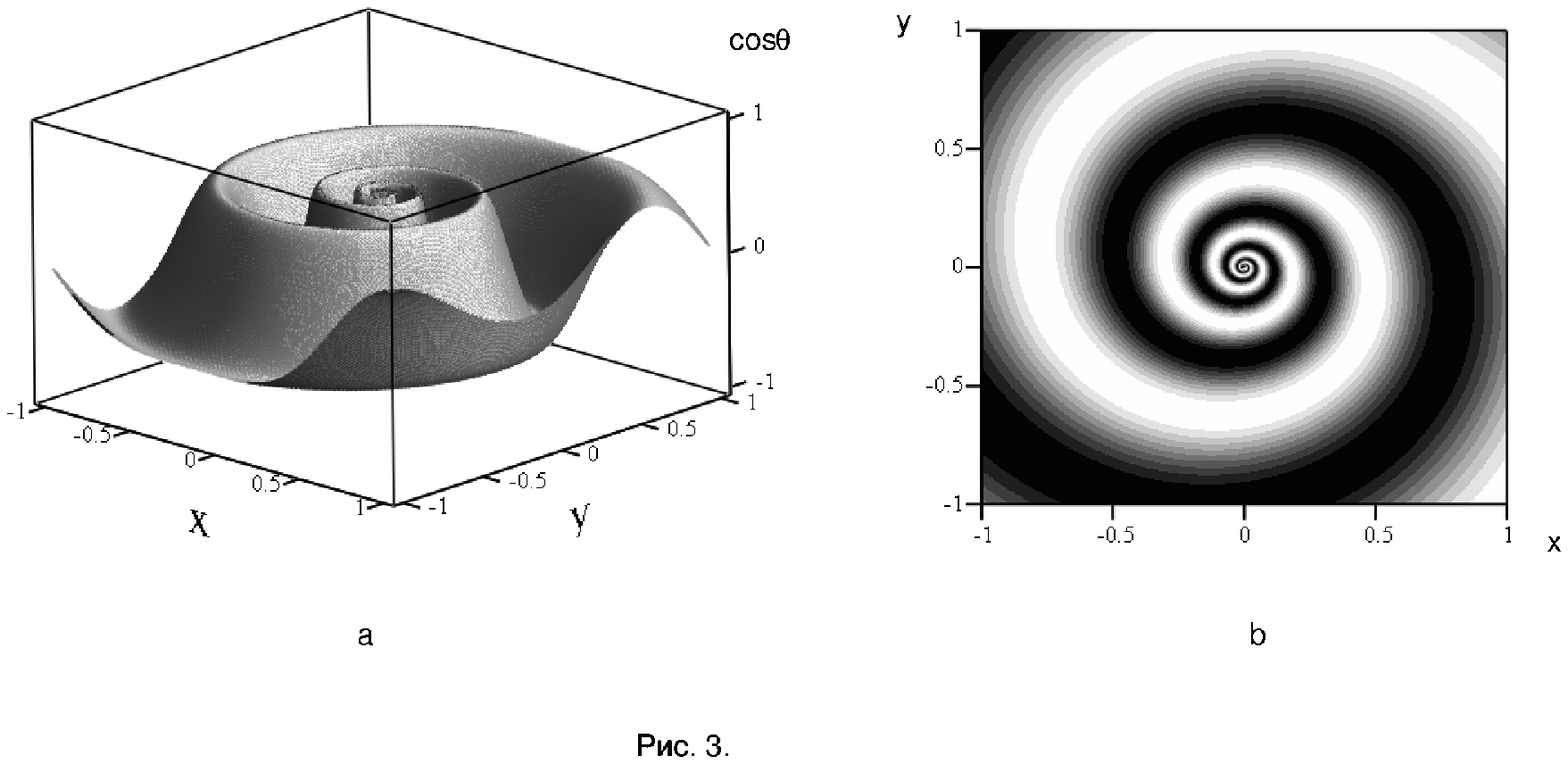}
\caption{Structure of the core of a one-turn spiral ($Q=1,N=1$, $c=0.5$, $k=0.15$).}
\label{F:fig-3}
\includegraphics[width=\columnwidth, viewport=10 110 560 340,clip]{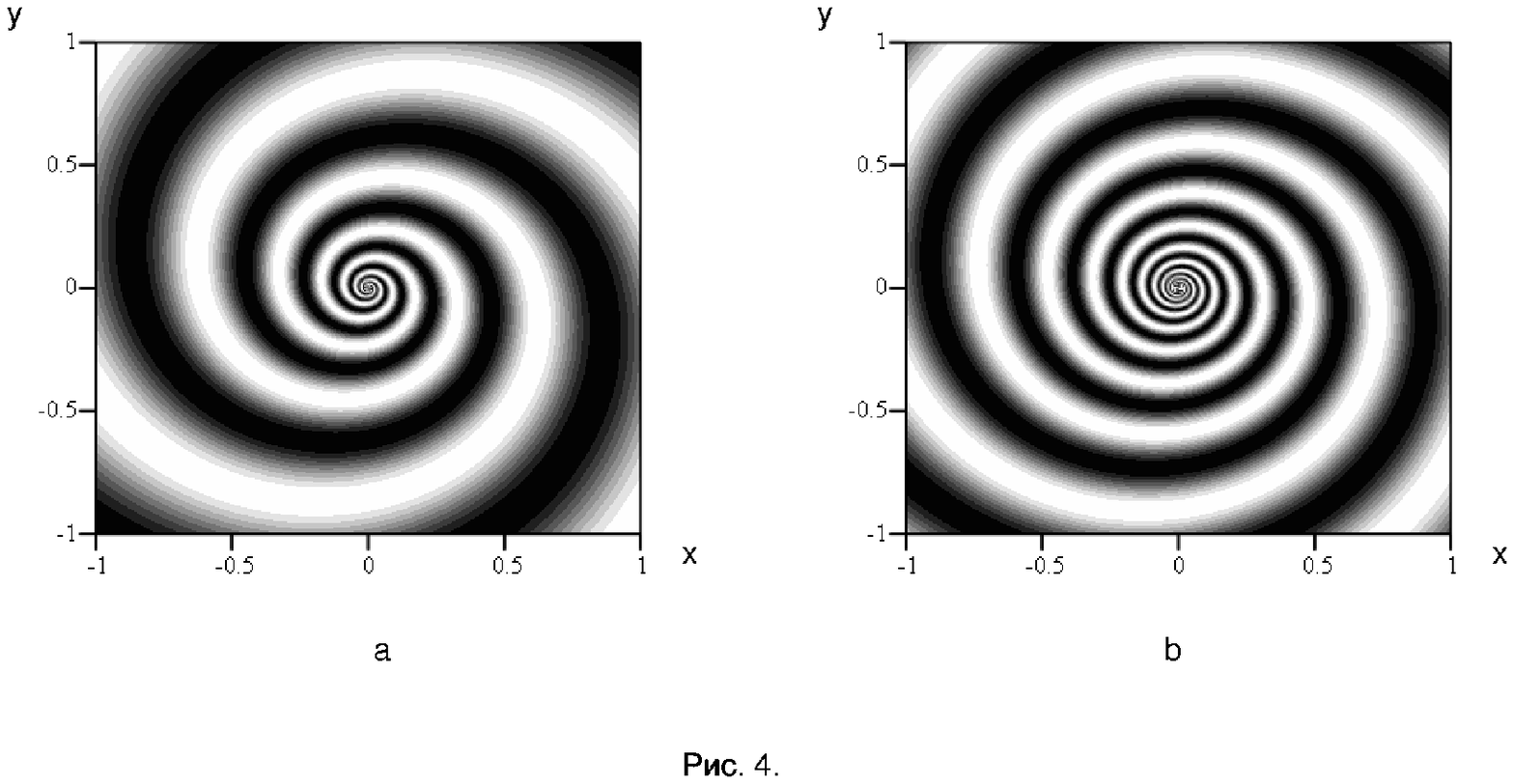}
\caption{Structure of the core of a one-turn spiral ($Q=1$, $N=1$, $c=0.5$);  $k=0.15$ (a), $k=0.1$ (b).}
\label{F:fig-4}
\end{figure}

\begin{figure}
\includegraphics[width=\columnwidth, viewport=40 110 560 340,clip]{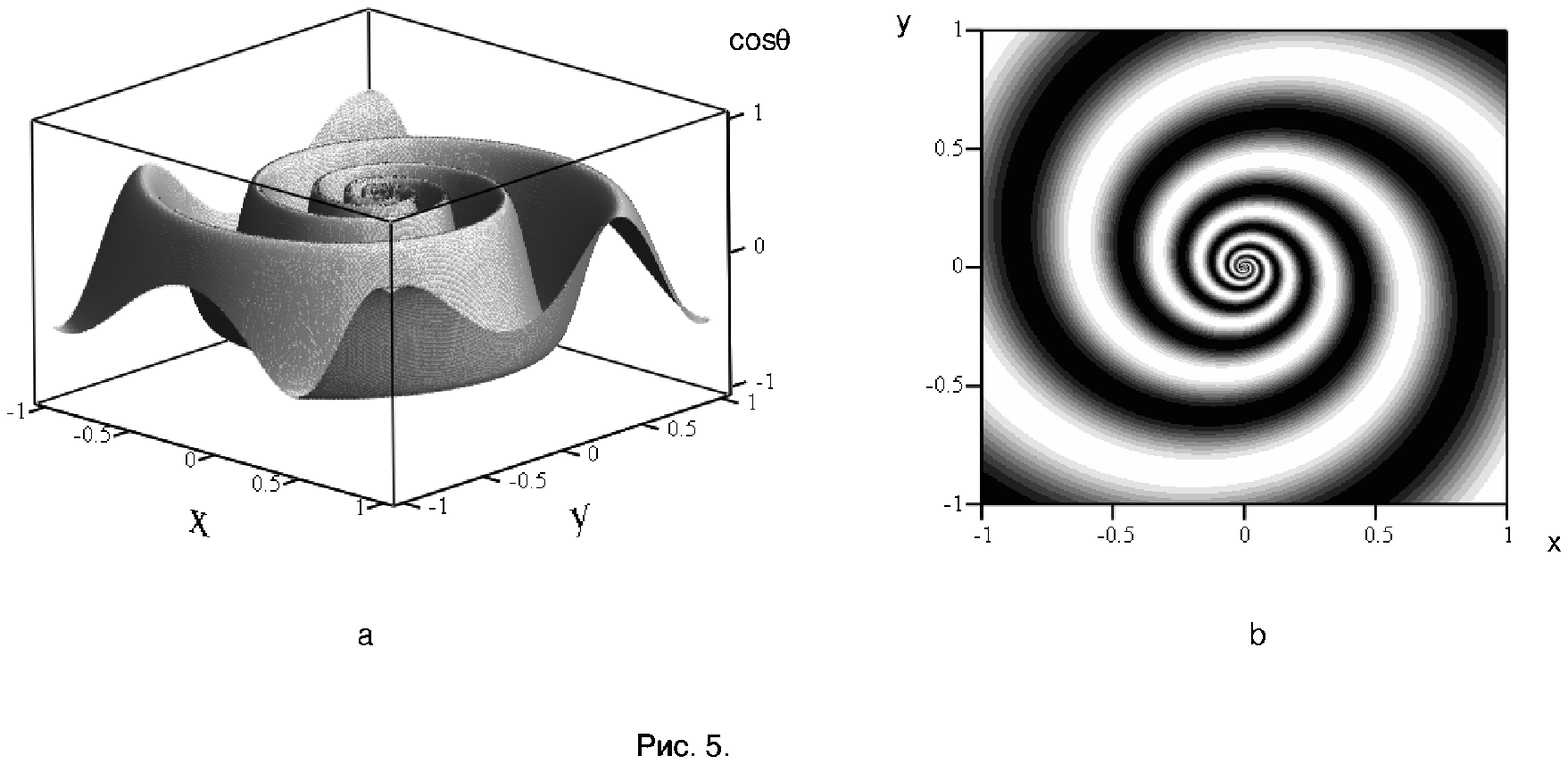}
\caption{Structure of the core of a two-turn spiral ($Q=1$, $N=2$, $c=0.5$, $k=0.15$).}
\label{F:fig-5}
\end{figure}

Finally, the relations (\ref{eq:eq12}),(\ref{eq:eq13}),(\ref{eq:eq14}),(\ref{eq:eq15}), (\ref{eq:eq17}), (\ref{eq:eq19}) and (\ref{eq:eq20}) describe
\emph{new types of magnetic structures} in ferromagnets in
the exchange approximation. For $n=1$ we obtain in a polar
coordinate system
\begin{eqnarray*}
cos(\theta)=c \cdot sn(\alpha \cdot ln(r) + \frac{2 K N}{\pi}\varphi, k),
\end{eqnarray*}
\begin{equation}
\Phi = \frac{k \alpha}{c}\varphi - \frac{2 K k N}{c \pi} ln(r) + \frac{ \sqrt{1-c^2}\sqrt{c^2-k^2}}{c} \int_0^{(\alpha\cdot ln(r) + \frac{2 K N}{\pi}\varphi)} \frac{dX}{(-1+c^2 sn(X,k)^2)},
\label{eq:eq21}
\end{equation}
where
\begin{eqnarray*}
\alpha=\frac{c Q}{k} + \frac{2 N \sqrt{1-c^2}\sqrt{c^2-k^2}}{k \pi}  \int_0^K \frac{dX}{(1 - c^2 sn(X,k)^2)}
\end{eqnarray*}
The parameter $c\leq1$ controls the "amplitude" with which
the spins leave the $xy$ plane. The structures (\ref{eq:eq21}) can be called
nutational, since the angle $\theta$ lies in the range $\theta_{max}\leq\theta<\pi-\theta_{max}$ with the maximum value $\theta_{max}=arccos(c)$. They include several types of structures. The case $N=0$ corresponds to a vortex magnetic "target" (Fig.\ref{F:fig-1}).
The distribution of the azimuthal
angle $\Phi$ is characteristic for vortex structures with
radial dependence, and the component $S_z$ has the form of
infinite concentric (with respect to the variable $r$) ring domains
forming an independent striped domain structure with
respect to the variable ln r. For $Q=0$ the azimuthal angle of
magnetization $\Phi$ depends on $r$, $\varphi$ and does not have a vortical
dependence. The component $S_z$ is a spiral structure, since
it is constant on the curves in the $(x, y)$ plane which are
logarithmic spirals (Fig.\ref{F:fig-2})
\begin{equation}
r=C exp \left(- \frac{2 K \varphi N}{\pi \alpha} \right).
\label{eq:eq22}
\end{equation}
An isolated magnetic defect with discrete parameters $N\neq0$ and $Q\neq0$
 is a spiral vortex with a vortical distribution of
the field $\Phi$ and a spiral structure for $S_z$. For $N=1$ the $S_z$
distribution consists of two domains have opposite directions
of magnetization and separated by two logarithmic spirals
(Fig.\ref{F:fig-3}). The width of spiral solitons (domain walls)
depends on $k$ and increases away from the center of the vortex.
Since $K=K(k)$ is a monotonically increasing function of the parameter
k, this parameter determines the degree of "twisting"
of the spiral (Fig.\ref{F:fig-4}). The chirality of the spiral (direction
of twist) is determined by the sign of the quantity $N/\alpha$. The
parameter $N$ determines the number of arms of the logarithmic
spiral. A plot of the field $S_z$ and the configuration of the
domains for a two-turn (two-arm) spiral are presented in Fig.\ref{F:fig-5}.
We shall show that the exchange interaction forms logarithmic
spirals $ln(r)\propto (\varphi - \varphi_0)$. Indeed, the equations (\ref{eq:eq7}) are
invariant under scale transformations $\overline{r}=r exp(\beta)$ and rotations
$\overline{\varphi}=\varphi+\gamma$ with the parameters $\gamma$ and $\beta$, respectively.
Consequently, the curve of constant values of $\theta(x,y)$ can be
invariant with respect to a single-parameter group of transformations
$\overline{r}=r exp(\beta)$, $\overline{\varphi}=\varphi+\rho \beta$ (spiral rotation group ~\cite{21})
with the parameter $\beta$. In our case $\rho=- \frac{\pi \alpha}{2 K N}$.

Direct calculations show that the energy density
$\frac{\nabla{\bf{S}}·\nabla{\bf{S}}}{2}$ of the structure (\ref{eq:eq21}) is proportional to $(\nabla a)^2$:
\begin{equation}
\frac{\nabla{\bf{S}}·\nabla{\bf{S}}}{2}=\frac{1}{2} (1+k^2-2 k^2 sn(a,k)^2)(\nabla a)^2.
\label{eq:eq23}
\end{equation}
To calculate the energy H of a spiral vortex we employ the
expansion~\cite{22}
\begin{eqnarray*}
sn(u,k)^2=\frac{1}{k^2 K^2} \left( K^2 - K E - 2 {\pi}^2 \sum_{n=1}^\infty \frac{n q^n}{1-q^{2n}} cos \left(\frac{2 n u \pi}{2 K}\right) \right).
\end{eqnarray*}
Here $E=E(k)$ is a complete elliptic integral of the second kind
and $q=exp(-\frac {K{'}\pi} {K})$ ($K{'}=K \sqrt{1-k^2}$). Then the energy of a
spiral vortex, just as that of other nonlocal structures of a
similar type (vortices and hydrodynamics, dislocations in the
crystal lattice), depends logarithmically on the size $L$ of the
system and the radius $d$ of the vortex core (of the order of the
lattice constant):
\begin{equation}
H=\frac{(2 E + (-1+k^2) K) ({\pi}^2{\alpha}^2+4 N^2 K^2)}{\pi K}ln\left(\frac{L}{d}\right)
\label{eq:eq24}
\end{equation}
where $N\neq0$.

Since a spiral is characterized by two integers $(N,Q)$,
spiral dipoles have structurally more diverse than vortices.
We shall examine as an example some types of vortex dipoles.
In contrast to many-instanton solutions, the energy of
multispiral configurations with $k\neq1$ depends on the distances
between the centers of the spiral vortices, which results
in their interaction.

\begin{figure}
\includegraphics[width=\columnwidth, viewport=40 125 560 340,clip]{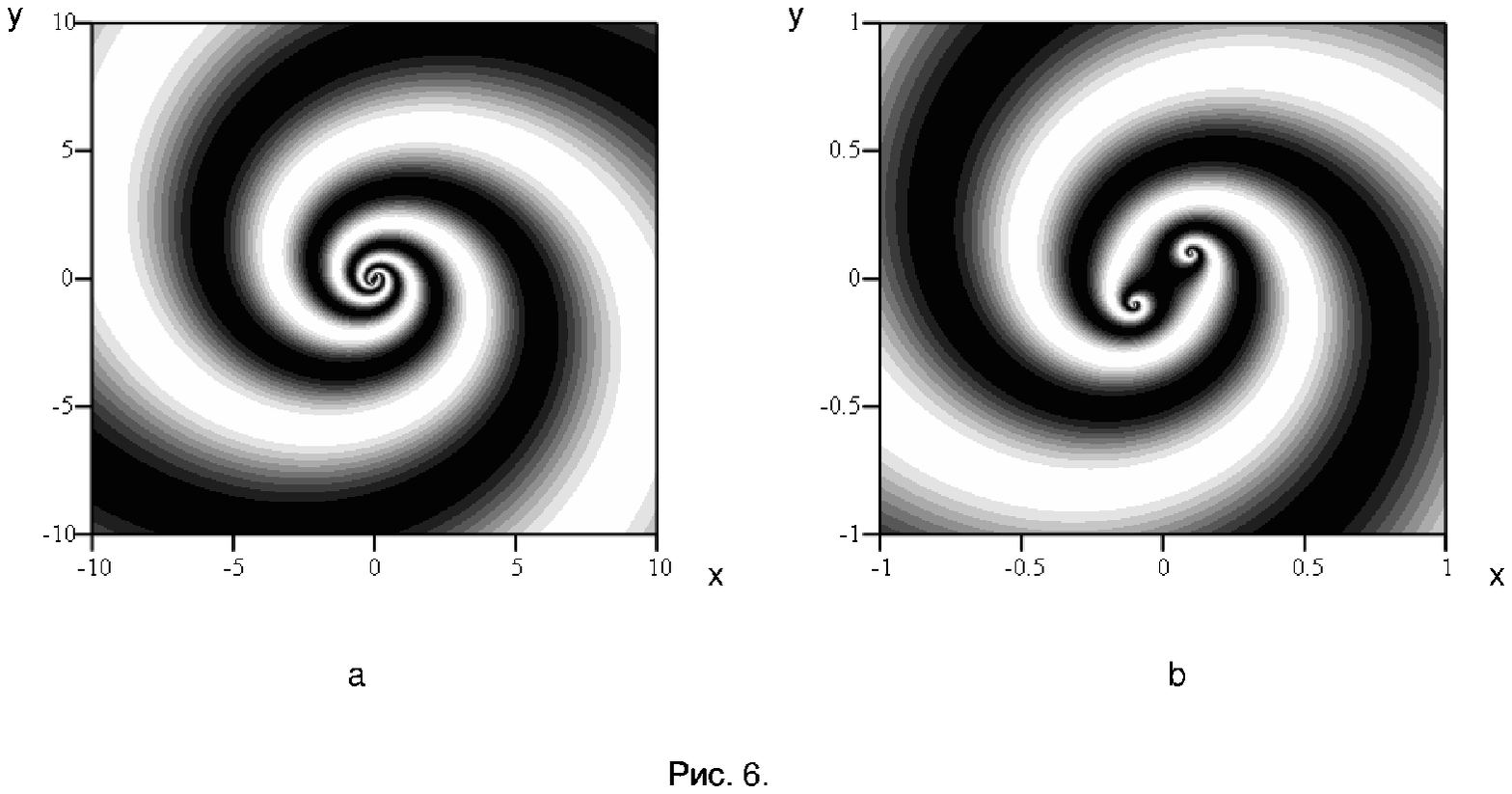}
\caption{Spiral dipole consisting of two one-turn spirals ($N=1,Q=0$ and $N=1,Q=0$), respectively, with $c=0.5$, $k=0.15$. Regions are shown on different scales: structure with large distances (a) and structure of the cores (b).}
\label{F:fig-6}
\includegraphics[width=\columnwidth, viewport=40 125 560 340,clip]{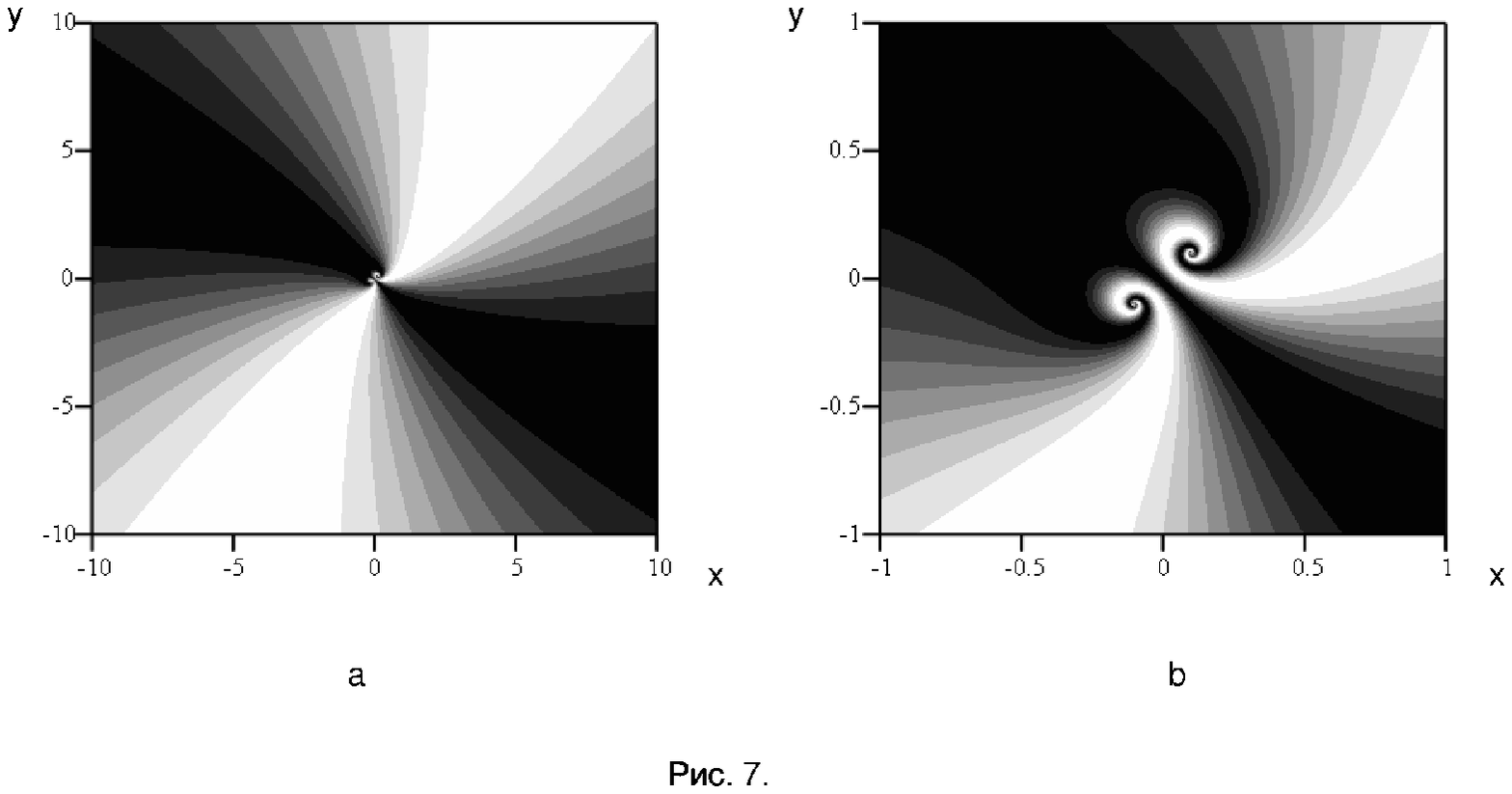}
\caption{Spiral dipole consisting of two one-turn spirals ($N=1,Q=0$ and $N=1,Q=-2$), respectively, with $c=0.5$, $k=0.15$.}
\label{F:fig-7}
\end{figure}

\begin{figure}
\includegraphics[width=\columnwidth, viewport=40 125 560 340,clip]{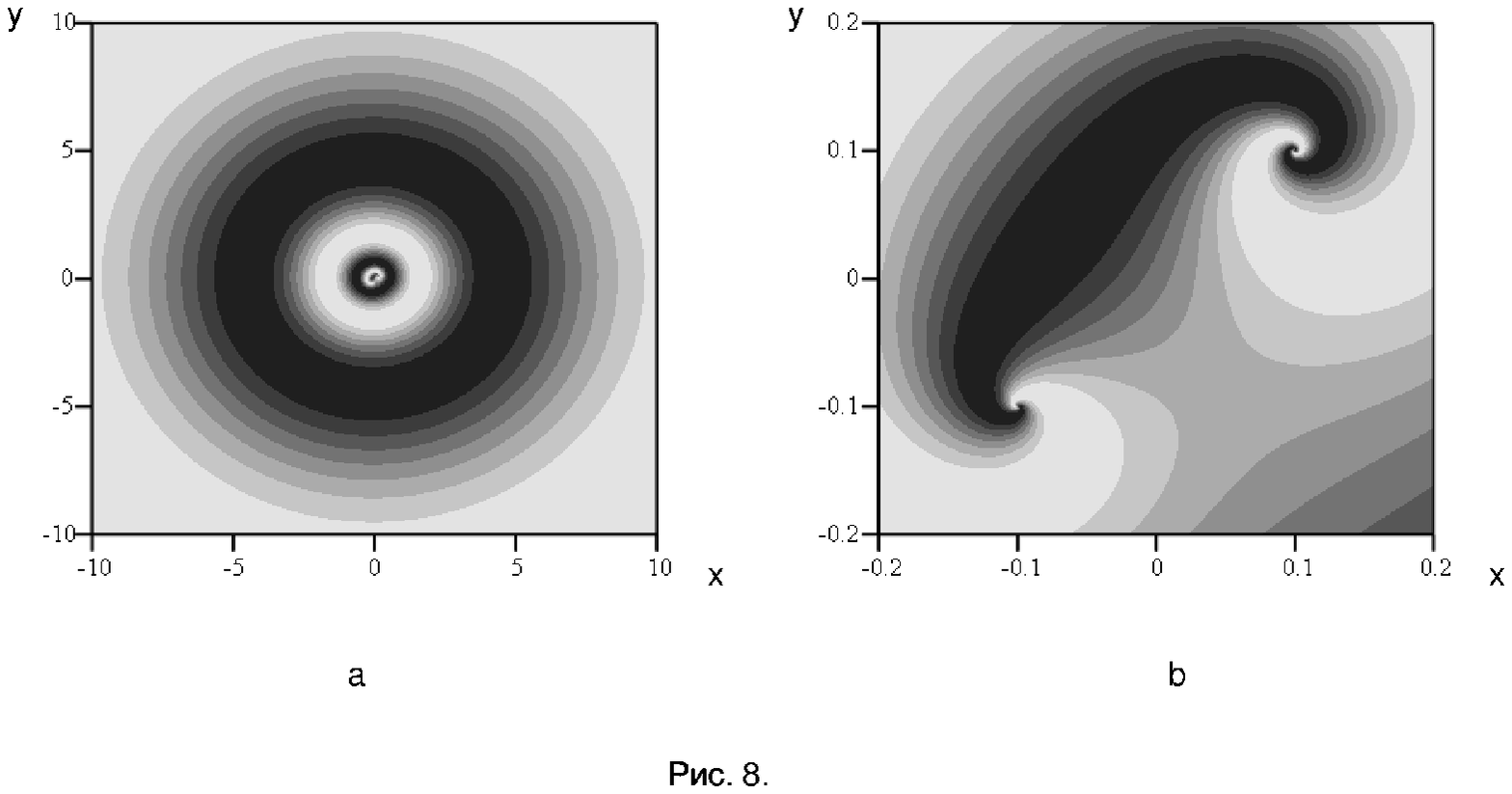}
\caption{Spiral dipole consisting of two one-turn spirals ($N=1,Q=0$ and $N=-1,Q=2$), respectively, with $c=0.4$, $k=0.25$.}
\label{F:fig-8}
\includegraphics[width=\columnwidth, viewport=40 125 560 340,clip]{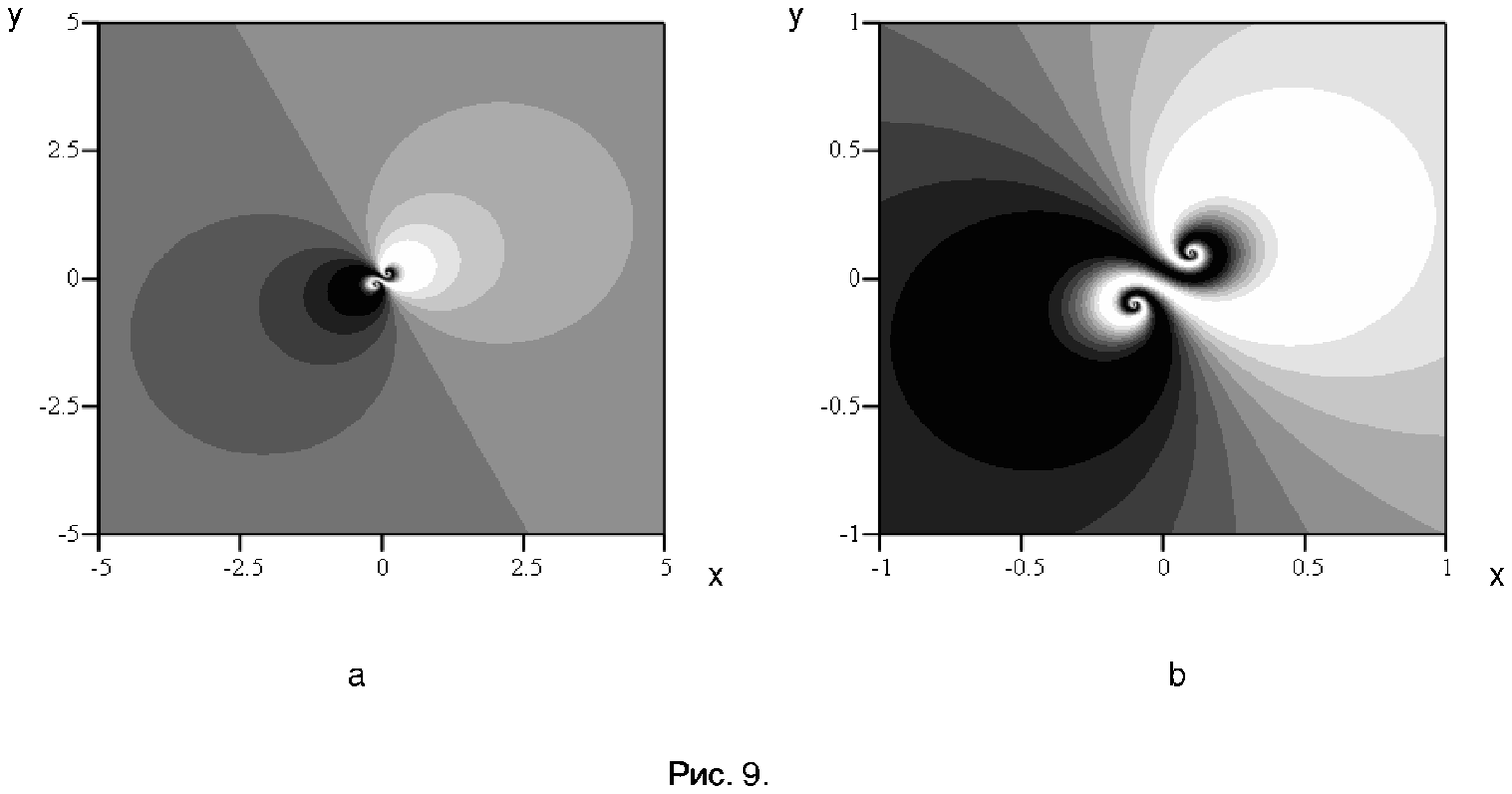}
\caption{Spiral dipole consisting of one-turn spirals ($N=1,Q=0$ and $N=-1,Q=0$), respectively, with $c=0.5$, $k=0.15$.}
\label{F:fig-9}
\end{figure}

As an example, we shall now examine certain types of
spiral dipoles consisting of vortex spirals with the numbers
$(N_1 ,Q_1)$ and $(N_2 ,Q_2)$. At large distances such a dipole transforms
into a definite spiral configuration with the numbers
$(N_1+N_2 ,Q_1+Q_2)$. A dipole consisting of two spiral structures
with the numbers $(1, 1)$ forms at large distances a two-turn
spiral (Fig.\ref{F:fig-6}), a dipole with $(1, 1)$ and $(1,.1)$ forms a $(2, 0)$
structure of a vortex-free spiral (Fig.\ref{F:fig-7}), and a dipole with $(1,1)$ and $(-1,1)$
forms a magnetic target structure (Fig.\ref{F:fig-8}).

The interaction of two vortices with the parameters
$(N,Q)$ and $(-N,-Q)$ is attractive. The corresponding solution
is localized and is displayed in Fig.\ref{F:fig-9}. The energy of
such a dipole does not depend on the size $L$ of the system
and at large distances the energy density is inversely proportional
to $r^4$. Since the activation energy is low, such spiral
dipoles can be generated by thermal fluctuations and contribute
to the thermodynamic properties of a system.

We shall discuss briefly the possibility of observing experimentally
the magnetic structures found in the present work.
The rapid advancement of the technology for growing
thin films has made it possible to produce artificially ordered
ASM alloys (artificially structured materials). As a result of
the influence of symmetry and low-dimension effects, new
phases can arise in such materials during growth of thin
films. Then a uniform state, conventionally considered to be
the ground state for a two-dimensional Heisenberg ferromagnet,
is simply impossible to obtain in practice if the magnetic
structure possesses a nonzero momentum or angular momentum.

We are deeply grateful to Professor A. S. Kovalev for
inviting us to participate in this issue of the journal, dedicated
to the memory of A. M. Kosevich.

\begin{thebibliography}{99}

\bibitem{1}
Nonlinearity in Condensed Media, edited by A. R. Bishop, R. Ecke, and S.
Gubernatis, Springer, Berlin (1993).

\bibitem{2}
Nonlinear Coherent Structures in Physics and Biology, edited by K. H.
Spatchek and F. G. Mertens, Plenum, New York (1994).

\bibitem{3}
Fluctuation Phenomena: Disorder and Nonlinearity, edited by A. R.
Bishop, S. Jimenez, and L. Vasquez, World Science, Singapore (1995).

\bibitem{4}
E. M. Lifshitz and L. P. Pitaevskii, Statistical Physics, Nauka, Moscow
(1978).

\bibitem{5}
I. S. Aranson and L. Kramer, Rev. Mod. Phys. \textbf{74}, 99 (2002).

\bibitem{6}
A. S. Kosevich, B. A. Ivanov, and A. S. Kovalev, Nonlinear Waves of
Magnetization. Dynamical and Topological Solitons, Naukova dumka,
Kiev (1983).

\bibitem{7}
T. Shinjo, T. Okuno, R. Hassdorf, K. Shigeto, and T. Ono, Science \textbf{289},
930 (2000); A. Leib, S. P. Li, V. Natali, and Y. Chen, J. Appl. Phys. \textbf{89}, 3892 (2001).

\bibitem{8}
P. S. Hagan, SIAM J. Appl. Math. \textbf{42}, 762 (1982).

\bibitem{9}
L. M. Pismen, Vortices in Nonlinear Fields, Clarendon Press, Oxford (1999).

\bibitem{10}
A. Yu. Loskutov and A. S. Mikhailov, Introduction to Synergetics, Nauka,
Moscow (1990).

\bibitem{11}
I. S. Aranson and L. Kramer, Rev. Mod. Phys. \textbf{74}, 99 (2002).

\bibitem{12}
G. S. Kandaurova, Usp. Fiz. Nauk \textbf{172}, 1165 (2002).

\bibitem{13}
F. V. Lisovskii and E. G. Mansvetova, Fiz. Tverd. Tela (Leningrad) \textbf{31} (1989).

\bibitem{14}
I. E. Dikshtein, F. V. Lisovskii, E. G. Mansvetova, and E. S. Chizik, Sov.
Phys. JETP \textbf{100}, 1606 (1991) [JETP \textbf{73}, 888 (1991)].

\bibitem{15}
M. C. Gross and H. C. Honenberg, Rev. Mod. Phys. \textbf{65}, 851 (1993).

\bibitem{16}
    \bibinfo{author}{F.~J. Nedelec},
    \bibinfo{author}{T. Surrey},
    \bibinfo{author}{A.~C. Maggs},
     \bibinfo{author}{S. Leibler},
    \bibinfo{journal}{Nature (London)} \textbf{\bibinfo{volume}{389}},
    \bibinfo{pages}{305} (\bibinfo{year}{1997}).

\bibitem{17}
    \bibinfo{author}{A.~B. Borisov},
    \bibinfo{journal}{ JETP Lett.} \textbf{\bibinfo{volume}{73}},
    \bibinfo{pages}{242} (\bibinfo{year}{2001}),\\
    \url{http://www.springerlink.com/content/r35626w82042hj27}.

\bibitem{18}
    \bibinfo{author}{A.~B. Borisov},
    \bibinfo{author}{I.~G. Bostrem},
    \bibinfo{author}{A.~S. Ovchinnikov},
    \bibinfo{journal}{ JETP Lett.} \textbf{\bibinfo{volume}{80}},
    \bibinfo{pages}{103} (\bibinfo{year}{2004}),\\
    \url{http://www.springerlink.com/content/c8u403786032133v}.

\bibitem{19}
    \bibinfo{author}{R.~Balakrishnan},
    \bibinfo{author}{A.~R. Bishop},
    \bibinfo{journal}{Phys. Rev. B} \textbf{\bibinfo{volume}{40}},
    \bibinfo{pages}{9194} (\bibinfo{year}{1989}),\\
    \url{http://prola.aps.org/abstract/PRB/v40/i13/p9194_1}.

\bibitem{20}
    \bibinfo{author}{A.~A. Belavin},
    \bibinfo{author}{A.~M. Polyakov},
    \bibinfo{journal}{ JETP Lett.} \textbf{\bibinfo{volume}{22}},
    \bibinfo{pages}{245} (\bibinfo{year}{1975}),\\
    \url{http://jetpletters.ac.ru/ps/1529/article_23383.shtml}.

\bibitem{21}
G. Birkhoff, Hydrodynamics [Russian translation], Izd. Inostr. Lit., Moscow (1954).

\bibitem{22}
E. T. Whittaker and J. H. Watson, A Course of  Modern Analysis [Russian Translation], Fizmatgiz, Moscow (1963).


\end{thebibliography}
\end{document}